\documentclass[prl,aps,superscriptaddress,twocolumn,longbibliography]{revtex4-2}

\usepackage{graphicx}
\usepackage{bm}
\usepackage{graphicx}
\usepackage{amsmath}
\usepackage{amssymb}
\usepackage{epstopdf}
\usepackage{color}

\begin{document}


\title{Anomalous Josephson Coupling and High-Harmonics in Non-Centrosymmetric Superconductors with $S$-wave Spin-Triplet Pairing}


\author{Yuri Fukaya}
\affiliation{CNR-SPIN, I-84084 Fisciano (Salerno), Italy, c/o Universit\'a di Salerno, I-84084 Fisciano (Salerno), Italy}

\author{Yukio Tanaka}
\affiliation{Department of Applied Physics, Nagoya University, Nagoya 464-8603, Japan}
\author{Paola Gentile}
\affiliation{CNR-SPIN, I-84084 Fisciano (Salerno), Italy, c/o Universit\'a di Salerno, I-84084 Fisciano (Salerno), Italy}

\author{Keiji Yada}
\affiliation{Department of Applied Physics, Nagoya University, Nagoya 464-8603, Japan}

\author{Mario Cuoco}
\affiliation{CNR-SPIN, I-84084 Fisciano (Salerno), Italy, c/o Universit\'a di Salerno, I-84084 Fisciano (Salerno), Italy}
\email[Corresponding author: ]{mario.cuoco@spin.cnr.it}

\begin{abstract}
We study the Josephson effects arising in junctions made of non-centrosymmetric superconductors with spin-triplet pairing having $s$-wave orbital-singlet symmetry. We demonstrate that the orbital dependent character of the spin-triplet order parameter determines its non-trivial texture in the momentum space due to the inversion symmetry breaking and spin-orbit interactions. The emergence of this pattern is responsible for the occurrence of an anomalous Josephson coupling 
and a dominance of high-harmonics in the current phase relation. Remarkably, due to the spin-orbital couplings, variations in the electronic structure across the heterostructure
can generally turn the ground state of the junction from 0- to a generic value of the Josephson phase, thus realizing the so-called $\varphi$-junction. 
Hallmarks of the resulting Josephson behavior, apart from non-standard current-phase relation, are provided by an unconventional temperature and magnetic field dependence of the critical current. These findings indicate the path for the design of superconducting orbitronics devices and 
account for several observed anomalies of the supercurrent in oxide interface superconductors.
\end{abstract}
\maketitle


\section{Introduction}

In the presence of conventional Cooper pairing the current-phase relation (CPR) of a superconductor-insulator-superconductor junction is given by $I_J=I_c \sin(\phi)$ \cite{Jos62}, with $I_c$ being the critical current and $\phi$ the phase difference.
While the sinusoidal shape is not always preserved \cite{Likharev1979}, as in extended superconductor-normal metal-superconductor junction or narrow ballistic weak links \cite{Bardeen1972,Ishii1970,Kulik1972,Kulik1978,Furusaki1991}, a vanishing supercurrent and non-degenerate minimum of the Josephson energy at $\phi=0$ are robust marks of the CPR in conventional Josephson junctions (JJ).
In this context, progress in materials science and nanofabrication have led to several physical cases with an unconventional CPR. 
Deviations from standard CPR indeed can manifest as a Josephson energy offset of a fractional flux quantum, leading to the so called $\varphi_0$-junction \cite{Buzding2008,Tanaka2009,Tanaka2014,Golubov2004,Tanaka2000,Asano2003,Grein2009,Eschrig2008,Silaev2017,Konschelle2019,Alidoust2021} that violates time-reversal symmetry. On the other hand, for an offset of half-integer flux quantum a $\pi$-junction \cite{HarlingenRMP1995} is realized. Apart from anomalous phase shifts, the energy of the Josephson junction can keep the symmetry of phase-inversion but exhibits a minimum at values of the phase which is different from 0 or $\pi$, setting out a $\varphi$-Josephson junction. Typical requirements to achieve a $\varphi$-junction are i) combination of 0- and $\pi$- Josephson couplings {{\cite{Goldobin2007,Goldobin2011,Yerin2014,Sickinger2012,guarcello2022}}}, ii) specific parameters range and geometric configuration of the junction \cite{TanakaKashi96,TanakaKashi97,Josephson2,KashiwayaTanaka2000RepProgPhys}, or iii) higher harmonics \cite{Goldobin2007}.

Observations of CPR in form of $\varphi_0$ or $\varphi$-junctions as well as 0-$\pi$ transitions have been reported in a variety of
devices based on InSb nanowires \cite{Szombati2016}, ferromagnetic heterostructures {{\cite{Sickinger2012,Ahmad2022}}}, superconducting spin valves \cite{Gincrich2016,DiBernardo2019},
and junctions with superconducting materials having non-trivial gap symmetry (e.g. iron picnitides \cite{Chen2010}, oxides interface \cite{Stornaiuolo2017,Sin21}, and cuprates \cite{Kir11,TanakaKashi96,TanakaKashi97}).
Here, competing 0- and $\pi$-Josephson channels lead to a vanishing first harmonic with a consequent dominant role of the second harmonic in determining the Josephson CPR. Having a CPR with non-negligible harmonics higher than the second one is however quite unusual and difficult to achieve without fine tuning. 

Recently, the combination of inversion symmetry breaking and multiple orbital degrees of freedom has emerged as an innovative route to tailor unconventional Josephson effects. This is mostly due to the expectation in acentric materials of a superconducting order parameter that goes beyond the canonical singlet-triplet mixed parity \cite{Frigeri2004}, as for the inter-band anti-phase pairing (e.g., $s_{+-}$ and $s_{+--}$)~\cite{Scheurer-2015-2,Mercaldo2020} or pure even-parity inter-orbital spin-triplet pairs~\cite{fukaya18}. 
Striking experimental evidences of anomalous Josephson effects and supercurrents have been indeed reported in noncentrosymmetric superconductors based on oxides interface \cite{Bal-2015,Stornaiuolo2017,Sin21} which have been ascribed to the occurrence of competing 0- and $\pi$-Josephson channels as well as to second harmonics in the CPR.
The microscopic origin of the observed effects is however not yet fully settled.

Motivated by this challenge and, in general, by the origin of anomalous Josephson effects in low-dimensional non-centrosymmetric superconductors (NCSs), we demonstrate how to achieve 0-, $\pi$- and $\varphi$- Josephson couplings together with robust high-harmonics in Josephson junction by exploiting multi-orbital degrees of freedom.
In spin-triplet superconductors the Cooper pairs are typically described by the so-called ${\bf d}$-vector whose components express the amplitude of the zero spin projection of the spin-triplet pairing states \cite{sigrist91}. Here, we find that the Josephson effects are a striking manifestation of the non-trivial orientations of the ${\bf d}$-vector in momentum space arising from the intertwining of its orbital dependent character with the inversion symmetry breaking and spin-orbit couplings. 
We demonstrate that $0$ to $\pi$ transitions and $\varphi$-Josephson phase can be generally realized and manipulated by varying the character of the electronic structure and the strength of inversion symmetry breaking interaction, thus being highly tunable through gating or electric field.
Hallmarks of a Josephson junction based on this type of NCSs are: i) CPR with $\varphi$-phase and dominant high-harmonics, ii) anomalous temperature dependence of the critical current with a linear upturn at low temperature and iii) maximum of the critical current at finite applied magnetic field in the Fraunhofer pattern. We discuss how this type of pairing can account for the recent observations on supercurrents behavior in oxides interface superconductors \cite{Stornaiuolo2017,Sin21}. 
\begin{figure*}
    \centering
    \includegraphics[width=15cm]{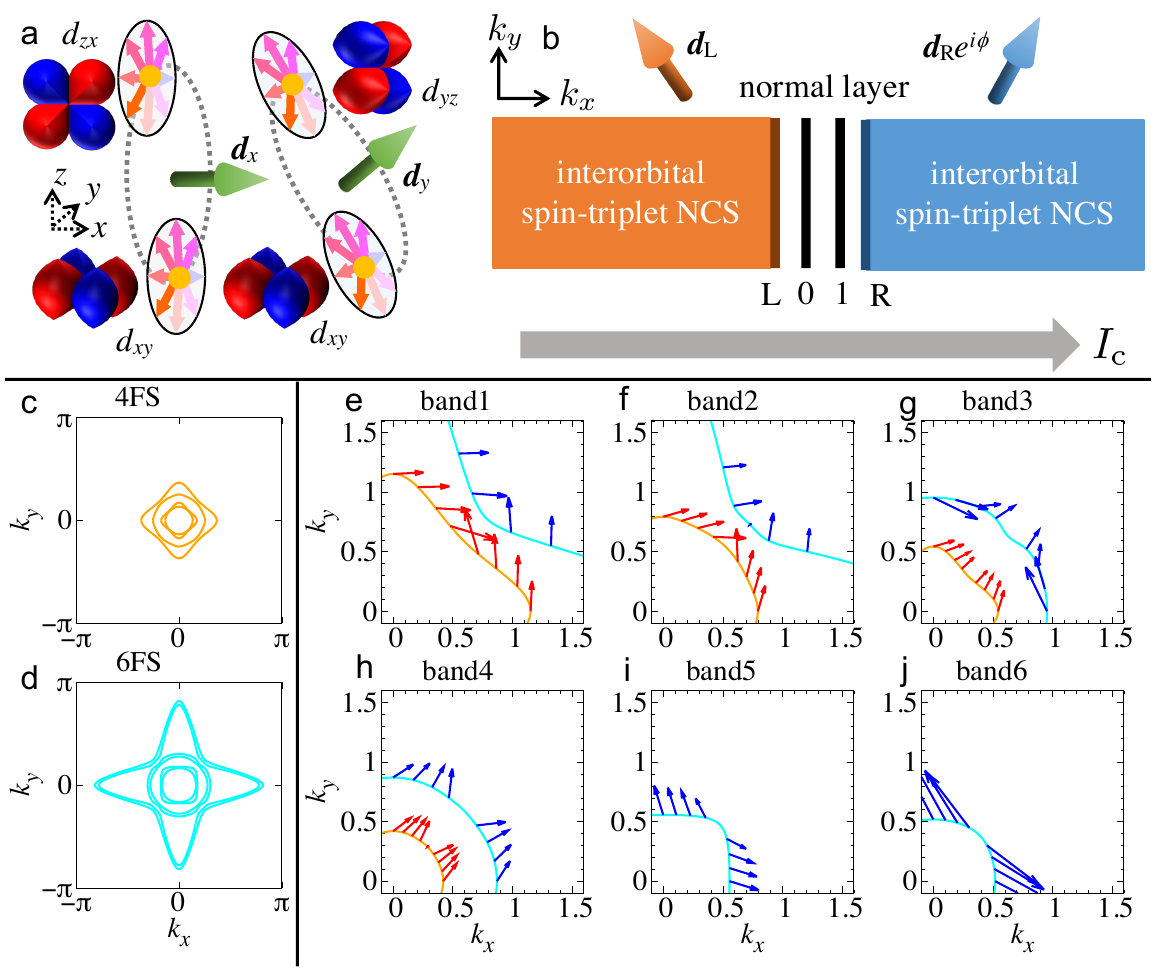}
    \caption{Spin-triplet Cooper pairs in momentum space along the Fermi line.
    \textbf{a} Interorbital spin-triplet Cooper pair and \textbf{b} Josephson junction. 
    Fermi lines in the normal state for different electronic parameters: \textbf{c} four Fermi lines at $(\lambda_\mathrm{SO}/t,\alpha_\mathrm{OR}/t,\mu/t)=(0.10,0.20,0.00)$ and \textbf{d} six Fermi lines configuration at $(\lambda_\mathrm{SO}/t,\alpha_\mathrm{OR}/t,\mu/t)=(0.10,0.20,0.35)$. 
    Spin-triplet \textbf{d}-vectors along the different Fermi lines are reported in the panels from \textbf{e} to \textbf{j} with orange and cyan indicating the Fermi contours corresponding to the configurations in \textbf{c} and \textbf{d}, respectively. The texture of the \textbf{d}-vector is significantly dependent on the band/orbital character and on the position in the momentum space.}
\end{figure*}%
\begin{figure*}
    \centering
    \includegraphics[width=15cm]{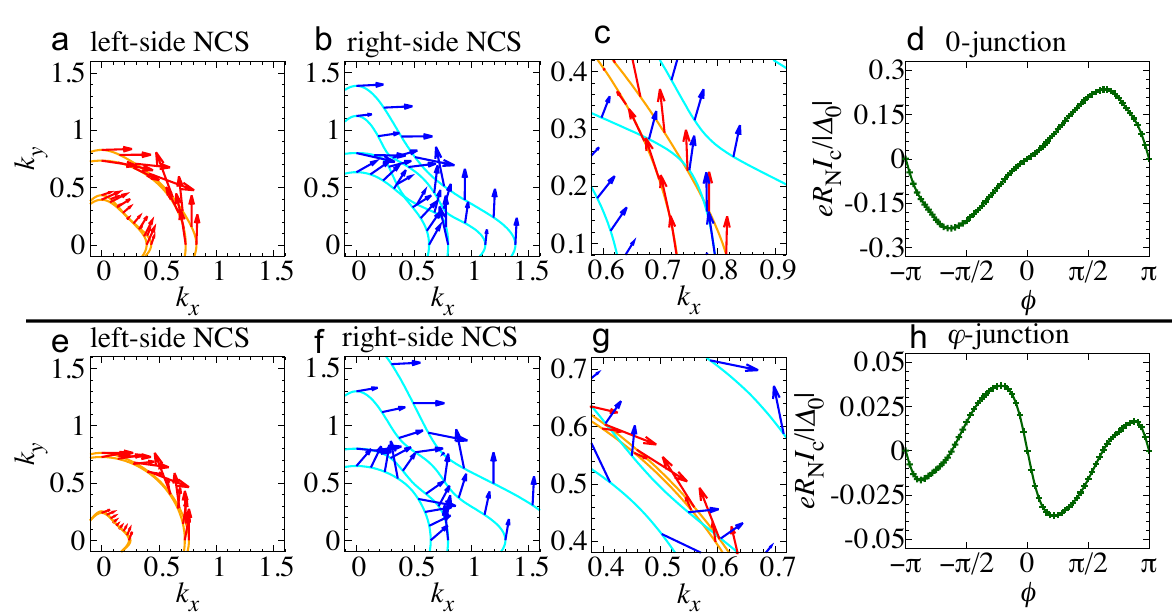}
    \caption{Current phase relation dependence on the orientation of the spin-triplet \textbf{d}-vector in the two sides of the junction.
    \textbf{a}, \textbf{b} and \textbf{c} indicate the \textbf{d}-vector orientation along the four Fermi lines on the left (L) and right (R) sides of the junction with a resulting current-phase relation \textbf{d} marked by a $0$-Josephson coupling. 
    \textbf{d}-vector along the four Fermi lines at \textbf{a} $(\lambda_\mathrm{SOL}/t,\alpha_\mathrm{ORL}/t,\mu_\mathrm{L}/t)=(0.06,0.05,0.00)$ and \textbf{b} $(\lambda_\mathrm{SOR}/t,\alpha_\mathrm{ORR}/t,\mu_\mathrm{L}/t)=(0.10,0.15,0.10)$ in the two NCS forming the junction. 
    \textbf{c} Zoom-in of the \textbf{d}-vector pattern for the electronic configurations in \textbf{a} and \textbf{b} around the overlapping points of the Fermi lines. 
    \textbf{d} Current phase relation evaluated by assuming the \textbf{a} and \textbf{b} electronic configurations on the two sides of the Josephson junction at $T/T_\mathrm{c}=0.025$, indicating a 0-type Josephson phase. 
    \textbf{d}-vector along the four Fermi lines for different electronic parameters: \textbf{e} left-side NCS $(\lambda_\mathrm{SOL}/t,\alpha_\mathrm{ORL}/t,\mu_\mathrm{L}/t)=(0.02,0.05,0.00)$ and \textbf{f} right-side NCS $(\lambda_\mathrm{SOR}/t,\alpha_\mathrm{ORR}/t,\mu_\mathrm{R}/t)=(0.20,0.30,0.10)$.
    \textbf{g} Zoom-in view of \textbf{e} and \textbf{f} nearby the overlapping points of the left- and right-side Fermi lines.
    \textbf{h} Current phase relation assuming the panels \textbf{e} and \textbf{f} electronic configurations on the two sides of the Josephson junction at $T/T_\mathrm{c}=0.025$. The computation of the current phase relations in \textbf{d} and \textbf{h} has been performed by means of the recursive Green's function method.}
\end{figure*}%


\section{Orbital-singlet spin-triplet pairs and $d$-vector profile along the Fermi lines} 

We consider a multi-orbital 2D electronic system with spin-triplet $s$-wave pairing.
In the normal state we have three bands arising from atomic orbitals spanning an $L=1$ angular momentum subspace, such as $d_a$ orbitals with $a=(yz,zx,xy)$. Here, we refer to $d$-orbitals localized at the site of a square lattice assuming a $C_{4v}$ point group symmetry. 
The breaking of mirror symmetry, in the plane of the junction, sets out a polar axis $z$ leading to an orbital Rashba interaction ($\alpha_\mathrm{OR}$) that couples the atomic angular momentum ${\bf L}$ with the crystal wave-vector ${\bf k}$ in the standard form $\sim [\hat{L}_x \sin(k_y)-\hat{L}_y \sin(k_x)$]\cite{park11,park12,Khalsa2013PRB,kim14,Mercaldo2020}. 
The atomic spin-orbit coupling ($\lambda_\mathrm{SO}$) expresses the interaction between the spin and angular momentum at each site. Taking the basis of the local creation operator of electrons of $d$-orbitals,
$\hat{C}^{\dagger}_{\bm{k}}=[c^{\dagger}_{yz,\uparrow\bm{k}},c^{\dagger}_{zx,\uparrow\bm{k}},c^{\dagger}_{xy,\uparrow\bm{k}},c^{\dagger}_{yz,\downarrow\bm{k}},c^{\dagger}_{zx,\downarrow\bm{k}},c^{\dagger}_{xy,\downarrow\bm{k}}]$, 
the Hamiltonian can be generally expressed as 
\begin{align}
    \hat{\mathcal{H}}=\sum_{\bm{k}}\hat{C}^{\dagger}_{\bm{k}}\hat{H}(\bm{k})\hat{C}_{\bm{k}} ,
\end{align}%
whose details of the matrix structure and the electronic dispersion with nearest-neighbor hoppings are provided in the Methods.  

In the superconducting state, the Bogoliubov-de Gennes (BdG) Hamiltonian is then directly constructed by including
the pair potential $\hat{\Delta}(\bm{k})$. 
In the present study, our focus is on the symmetry allowed local ($s$-wave) spin-triplet pairing with orbital-singlet character and B$_1$ symmetry in the $C_{4v}$ group. 
This type of pairing is energetically favorable when considering that inter-orbital interactions are dominant with respect to the intra-orbital ones \cite{fukaya18}. The  B$_1$ order parameter is described \cite{fukaya18,fukaya19,Fuk20} by a $\bm{k}$-independent \textbf{d}-vector with $d_x$ ($d_y$) components corresponding to local electron pairs between $\{d_{zx},d_{xy}\}$ ($\{d_{zy},d_{xy}\}$) orbitals, respectively. 

By introducing orbital indices to label the ${\bf d}$-vector, we have that $d^{(xy,zx)}_{x}=d^{(xy,yz)}_{y}$.
A schematic illustration of the spin and orbital structure of the Cooper pair is shown in Fig.~1a.
This type of pairing exhibits nodal points along the diagonal of the Brillouin zone ([110] direction) with nonzero topological number that is due to the chiral symmetry of the BdG Hamiltonian \cite{Sato2011PRB,Yada2011PRB,Brydon2011PRB,Mercaldo16,fukaya18,fukaya19,Fuk20}. Before analysing the Josephson effects in the junction formed by interfacing NCS [Fig.~1b], it is useful to consider the structure of the spin-triplet pairing in momentum space along the Fermi lines. This is evaluated by considering the anomalous components of the Green's function (see Methods for details).
To this aim we choose two representative configurations whose electron density yields four [Fig.~1c] and six [Fig.~1d] Fermi lines around the $\Gamma$ point. The inner Fermi lines are more isotropic, while the outer ones exhibit a more pronounced anisotropy. 
For the examined 2D tetragonal configuration, the $xy$ orbital at $k=0$ is lower in energy as compared to the $(zx,yz)$ bands. 
We start by observing how the orientation of the ${\bf d}$-vector at a given $\bm{k}$-point is modified along the Fermi lines. For those bands having large Fermi momentum, i.e. bands 1 and 2 in Fig.~1e and Fig.~1f, we have obtained that the ${\bf d}$-vector is mostly pointing along the $x$ ($y$) direction for crystal wave vectors above (below) the diagonal. This is because the electronic configurations at the Fermi level have dominant $xy$ character and mixing with $zx$ and $yz$ states through the spin-orbit and the orbital Rashba couplings. When we consider the bands closer to the center of the Brillouin zone [Figs.~1g-j], we observe that the ${\bf d}$-vector exhibits a completely different pattern which is marked by an orientation that is mostly along the [110] direction [Figs.~1g,h] or perpendicular to it [Figs.~1i,j]. Such behavior is mostly due to the dominant $(zx,yz)$ character of the electronic states and the fact that the B$_1$ pairing does not involve a direct coupling between such orbital states, thus implying an equal weight for $d_x$ and $d_y$ components. 
The results in Figs.~1c-j provide evidence of the Cooper pairs along the Fermi lines having a spin-triplet configuration not uniform in orientation and amplitude that strongly depends on the orbital character of the corresponding electronic states.

\section{Misalignment of $d$-vectors across the interface and 0-$\varphi$ Josephson coupling} 

Let us now consider the Josephson current in the NCS-NCS junction. 
For the computation of the Josephson current, we adopt the recursive Green's function method \cite{Lee_FisherPRL1981,LDOSUmerski} assuming two semi-infinite superconductors and setting the pair potential in the Josephson junction as $\hat{\Delta}_\mathrm{L}=\hat{\Delta}_\mathrm{B1}$ and $\hat{\Delta}_\mathrm{R}=\hat{\Delta}_\mathrm{B1}e^{i\phi}$,
with $\phi$ being the phase difference between the superconductors forming the junction [Fig.~1b].
Below the critical temperature $T_\mathrm{c}$,
Josephson current $I_\mathrm{c}(\phi)$ at the temperature $T$ is then determined \cite{KawaiPRB,Fuk20} as
\begin{align}
I_\mathrm{c}(\phi)=-\frac{iek_\mathrm{B}T}{\hbar}
    \int^{\pi}_{-\pi}dk_{y}\mathrm{Tr}'
    \sum_{i\varepsilon_{n}}[&\tilde{t}^{\dagger}_\mathrm{N}\hat{G}_{01}(k_y,i\varepsilon_{n},\phi)\notag\\
    -&\tilde{t}_\mathrm{N}\hat{G}_{10}(k_y,i\varepsilon_{n},\phi)],
    \label{JJ}
\end{align}%
with $\tilde{t}_\mathrm{N}$ being the nearest-neighbor hopping matrix in the normal layer, the non-local Green's functions $\hat{G}_{01}(k_y,i\varepsilon_{n},\phi)$ and $\hat{G}_{10}(k_y,i\varepsilon_{n},\phi)$ [Fig.~1b and Methods], and the fermionic Matsubara frequency $i\varepsilon_{n}=i(2n+1)\pi k_\mathrm{B}T$.
Here, $\mathrm{Tr}'$ means that the trace is only in the electron space.

Having demonstrated that the ${\bf d}$-vector has a non-trivial texture along the Fermi lines, we expect that the relative orientation of ${\bf d}$-vectors in the two sides of the junction has a key role in setting out the Josephson effect. To this aim we employ two representative ${\bf d}$-vectors configurations with a distribution of misalignment angles $\gamma $ that can be $\sim 0$ or close to $\pi/2$. The ${\bf d}$-vector along the four Fermi lines of the left (L) NCS [Fig. 2a] and right (R) NCS [Fig. 2b] is evaluated by considering different values for the electron filling via $\mu$, and the spin-orbital couplings through $\alpha_{\mathrm{OR}}$ and $\lambda_{\mathrm{SO}}$ (Supplementary Figures 1,2,3). Here, we focus on the region of momentum space where the left- and right-side Fermi lines mostly overlap. We start considering electronic configurations that result into ${\bf d}$-vectors that are about collinear on the two sides of the junction [Fig. 2c]. Hence, taking into account those spin-triplet tunneling processes, we find that the derivative of Josephson current is positive at low $\phi$ and the CPR has a standard profile with maximum at about $\phi \sim \pi/2$ [Fig. 2d]. 
Next, as reported in Figs.~2e-g, the texture of ${\bf d}$-vector in the region nearby the diagonal of the Brillouin zone at the crossings of the left- and right-side Fermi lines indicate that the misalignment angle is about $\pi/2$. On the other hand, for momentum along the [100] or [010] directions, the ${\bf d}$-vectors are mostly aligned. 
In this case, the resulting Josephson current [Fig.~2h] is marked by a change of sign in the derivative of Josephson current at low $\phi$ together with vanishing critical current for $\phi$ different from the time-reversal points at $\phi=0$ or $\pi$. This implies that the ground state realizes the so called $\varphi$-Josephson configuration. By tuning the amplitude of the orbital Rashba coupling or of the spin-orbit interaction it is possible to induce $0$ to $\pi$ Josephson phases and observe CPR dominated by harmonics higher than the second one close to the critical points (Figs.\ S2, S3 in the Supplementary Information).

To understand these Josephson effects, we recall that as shown in Ref. \cite{Asano2006-d} for a specific geometrical design of the junction, the Josephson current $J(\phi)$ between spin-triplet superconductors marked by ${\bf d}$-vectors that are misaligned by an angle $\gamma$ is expressed as
\begin{align}
J(\phi)
    \propto\frac{\sin{(\phi+\gamma)}}{\sqrt{1-|Z_{t}|^2\sin^2\left(\frac{\phi+\gamma}{2}\right)}}
    +\frac{\sin{(\phi-\gamma)}}{\sqrt{1-|Z_{t}|^2\sin^2\left(\frac{\phi-\gamma}{2}\right)}}
\end{align}
with $\phi$ being the applied phase difference and $Z_{t}$ the transmission amplitude across the junction. This expression indicates that in the tunneling process the spin-triplet pairs with opposite spin polarization undergo an antiphase shift to keep the time-reversal symmetry with an amount that is proportional to the misalignment angle $\gamma$. Here, it is immediate to deduce that for a value of $\gamma$ that is about $\pi/2$, the current $J(\phi)$ at small phase difference can change sign, thus turning the Josephson phase behavior from 0 to $\pi$. Since for the examined NCS-NCS junction the ${\bf d}$-vector has a variation of the orientation in the momentum space, it is useful to consider the Josephson behavior resulting from the superposition of different misaligned ${\bf d}$-vectors. 
Taking the representative case of a pair of Josephson channels with different configurations of ${\bf d}$-vector misalignment, as in Figs.~3a,b, one can find that when the angles $\gamma$ and $\gamma_1$ are both close to $\pi/2$, the resulting current is marked by non-vanishing and comparable amplitude of the first four harmonics (i.e $I_1$,$I_2$,$I_3$,$I_4$). In this case, the CPR has a profile that yields a $\varphi$-Josephson coupling.

\begin{figure*}
    \centering
    \includegraphics[width=15cm]{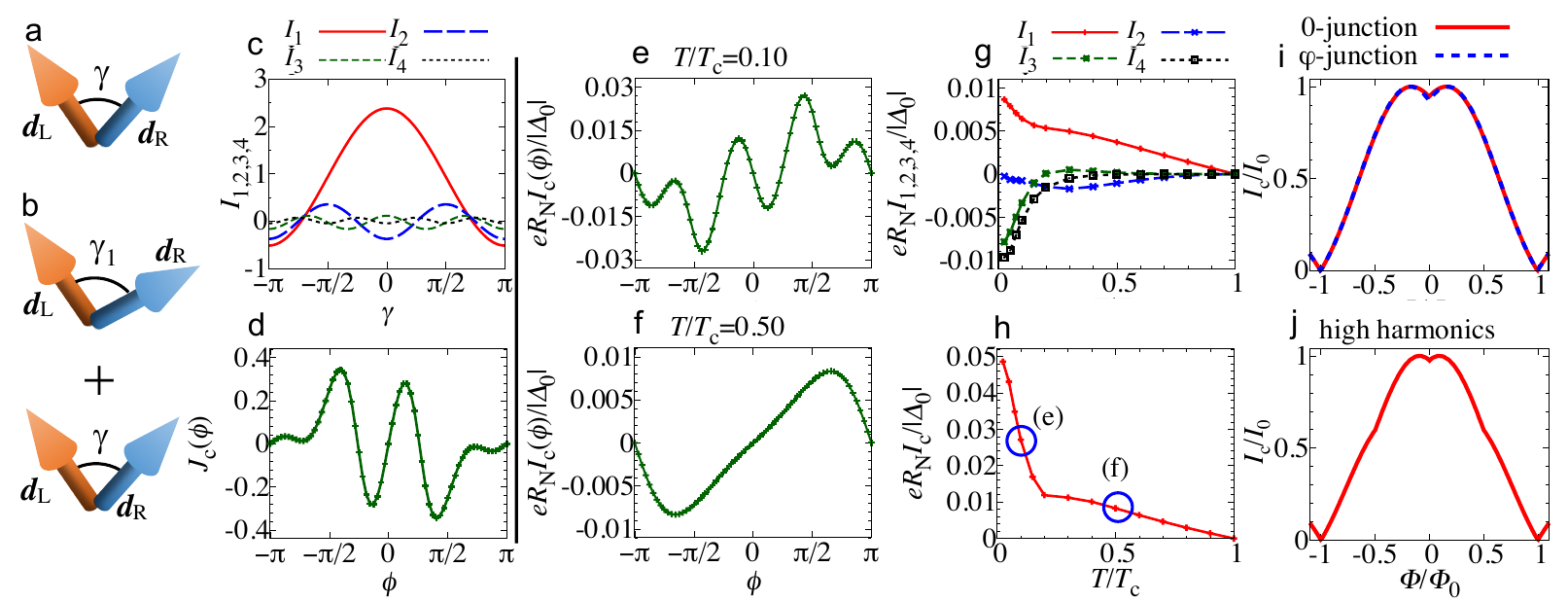}
    \caption{Misalignement of \textbf{d}-vectors, high-harmonics in the current phase relation and magnetic field dependence of the critical current.
    \textbf{a} Sketch of the \textbf{d}-vectors on the left (L) and right (R) sides of the junction with a misaligned angle $\gamma$.  
    \textbf{b} Schematic illustration of two Josephson channels with misaligned angles $\gamma$ and $\gamma_1$ for the \textbf{d}-vectors. 
    \textbf{c} The behavior of the first four harmonics amplitude ($I_n$ with $n=1,...,4$) of the Josephson current in the case of two Josephson channels, characterized by a varying misalignment angle ($\gamma$) and a given one ($\gamma_1=0.24\pi$) for the corresponding \textbf{d}-vectors. The transmission amplitudes are $Z_{t1}=0.85$ and $Z_t=0.95$.
    (d) Current phase relation for two Josephson channels with $Z_{t1}=0.85$, $Z_t=0.95$, $\gamma_1=0.24\pi$, and $\gamma=0.74\pi$.
    Current phase relation for the NCS-NCS junction at different temperatures evaluated by means of the recursive Green's function: \textbf{e} $T/T_\mathrm{c}=0.10$ and  \textbf{f} $T/T_\mathrm{c}=0.50$.
    \textbf{g} Temperature dependence of the $n=1,2,3,4$ Josephson harmonics and
    \textbf{h} the maximum amplitude of the Josephson current. 
    In \textbf{e}-\textbf{h}, we set parameters as $T_\mathrm{c}/t=1.0\times 10^{-5}$, $(\lambda_\mathrm{SOL}/t,\alpha_\mathrm{ORL}/t,\mu_\mathrm{L}/t)=(0.10,0.20,0.00)$, and $(\lambda_\mathrm{SOR}/t,\alpha_\mathrm{ORR}/t,\mu_\mathrm{R}/t)=(0.10,0.20,0.35)$.
    Magnetic field dependence of the critical current (Fraunhofer pattern) with an applied magnetic flux $\Phi$ threading the NCS-NCS junction for \textbf{i} 0- (red line) and $\varphi$-Josephson coupling (blue dotted line) as for the CPR in Fig.\ 2 \textbf{d} and \textbf{h}. \textbf{j} Fraunhofer pattern for a CPR dominated by high harmonics components in the Josephson current as in Fig.\ 3 \textbf{e}. 
}
\end{figure*}


Another peculiar hallmark of the Josephson current in the examined NCS-NCS junction is represented by the occurrence of high-harmonics in the CPR with nonvanishing and comparable amplitude. We find that this behavior can be also ascribed to the nontrivial misalignment of the ${\bf d}$-vectors across the junction and the superposition of different configurations along the Fermi lines.
Indeed, let us consider the Josephson current obtained by summing up only two channels [Fig.~3b], one with a given misalignment angle ($\gamma_1$) and the other one ($\gamma$) to vary in the range $[-\pi,\pi]$ [Fig.~3c]. Within this effective description, we find that high-harmonics generally occur in the CPR [Fig.~3d]. The CPR has an anomalous profile with a positive derivative at low $\phi$ and multiple oscillations due to the competing high harmonics. 
Since the ${\bf d}$-vector texture is not homogeneous along the Fermi lines, it is natural to expect that various channels with inequivalent mismatch angles and transmission amplitudes cooperate to build up such behavior for the supercurrent. In Figs.~3e,f we provide evidence of the CPR for the NCS-NCS junction, whose profile is significantly marked by high-harmonics at low temperature which in turn tend to get suppressed by increasing the temperature towards the transition into the normal state. 
Finally, we evaluate the response to an applied magnetic field that acts to modulate the critical current amplitude (details in the Supplementary Information). We find that the resulting Fraunhofer pattern [Figs.~3i,j], as a consequence of the intrinsic competing $0$ and $\pi$ Josephson couplings, does not have a maximum at zero magnetic flux, as expected in conventional spin-singlet JJ junctions. Interestingly, even for a CPR with a 0-Josephson coupling, as in Fig.~3i, the critical current has a local minimum rather than a maximum at zero applied magnetic field [Figs.~3i,j]. This demonstrates that the intrinsic tendency of having competing Josephson channels in the case of NCS superconductors with $s$-wave orbital-singlet and spin-triplet pairing has clearcut and detectable signatures in the magnetic field response. 



\section{Discussion}

We now discuss the impact of our findings in SrTiO$_3$ based hetero-structures, such as LaAlO$_3$/SrTiO$_3$ (LAO-STO)~\cite{Ohtomo-2004, Reyren-2007}.
The LAO-STO is an ideal 2D electron system with non-centrosymmetric multi-orbital superconductivity~\cite{fukaya18, Scheurer-2015-2, Fernandes-2013} exhibiting a remarkable control of the superconducting critical temperature by electrostatic gating ~\cite{Caviglia-2008,Thierschmann-2018,Hurand-2015} together with Rashba spin-orbit coupling~\cite{Caviglia2010PRL,Ben_Shalom2010PRL} and the occupation of the Ti 3d orbitals ($d_{xy}$,$d_{zx}$,$d_{yz}$)~\cite{Joshua-2012, Herranz-2015}.
The superconducting phase exhibits several anomalous properties that cannot be easily addressed within a conventional spin-singlet scenario. Clearcut unconventional observations are provided by the superconducting gap suppression nearby the Lifshitz transition~\cite{Singh-2019,Trevisan-2018}, the anomalous magnetic field dependence of critical current in weak links~\cite{Bal-2015,Stornaiuolo2017} and uniform nanowires~\cite{Kalaboukhov-2017}, and several in-gap bound states probed by tunneling spectroscopy~\cite{Kuerten-2017}. The significant role of inhomogeneities also poses 
fundamental questions on the nature of the superconducting state in LAO-STO interface, excluding $p$-wave spin-triplet pairing as a candidate, and pointing to an even-parity ($s$-wave) multi-band superconductivity which is robust to disorder.

Recently, superconducting transport measurements in nano-devices \cite{Stornaiuolo2017,Sin21} have provided direct and significant evidences of an orbital dependent unconventional pairing.
The central experimental findings demonstrate: i) an anomalous enhancement of the critical current at weak applied magnetic field, ii) an asymmetric response with respect to the magnetic field direction, and iii) the supercurrent anomalies are gate dependent, getting pronounced when reaching the Lifshitz transition at the onset of the occupation of the ($d_{zx}$,$d_{yz}$) bands.
In this context, our results can account for the central observations of the superconductng transport measurements. Our analysis is based on planar 2D junctions as for the experimental configuration and the unveiled Josephson effect, due to the inter-orbital spin-triplet pairing, leads to $\pi$-phase and high-harmonics in the Josephson current. As demonstrated by the analysis of the Fraunhofer pattern, the combination of $\pi$-phase and high-harmonics provides the enhancement of the critical current at weak applied magnetic field and we find it to be robust with respect to variations of the electronic parameters in the two sides of the junction. This result can be particularly relevant for the LAO-STO where the gating is related to an inhomogeneous distribution of the electron density in the 2D electron gas and of the inversion symmetry breaking at the interface. Concerning the asymmetric response with respect to the magnetic field direction, our results indicate that the combination of $\pi$-phase and high-harmonics is not sufficient to yield the effect. One has to consider it in a spatially inhomogeneous array of Josephson junctions \cite{Sin21}. Hence, we expect that the observed asymmetry of the Fraunhofer pattern with respect to the magnetic field can be dependent on the spatially homogeneity of the superconductor in terms of granularity and inclusion of nanometric sized islands.

Our findings highlight the relevant role of $s$-wave inter-orbital spin-triplet pairing in noncentrosymmetric junctions for achieving competing $0$- and $\pi$-Josephson couplings together with high-harmonics in the current phase relation.
Differently to single-band superconductors, the resulting CPR is tied to the character of the electronic structure and thus can be potentially tuned in a controlled way by changing the electron density and the Rashba coupling through gating. This aspect has a direct impact on the transport properties of LAO-STO junctions and points to distinct design of Josephson devices. We also point out that the distribution of the spin moment of the Cooper pairs in momentum space is a key quantum resource for the generation of CPR with dominance of high-harmonics. Remarkably, in all these configurations the magnetic field response reveals a Fraunhofer pattern with a minimum of the critical current at vanishing field. {{While nonstandard magneto-electric effects \cite{Brydon2009,Mercaldo2019a,Kotetes2019,Sakurai2020} are based on the magnetic field tunability of the \textbf{d}-vector orientation \cite{Gentile2013,Brydon2009,Mercaldo16}, our findings highlight the role of the orbital rather than the spin degree of freedom.}}
Hence, in perspective, due to the demonstrated intrinsic orbital tunability of the \textbf{d}-vector texture by orbital Rashba coupling, our findings set out innovative routes to design orbitally driven magneto-electric effects and Josephson devices for superconducting orbitronics with orbital control of the supercurrent.

\section{Acknowledgements}

This research has received funding by ERA-NET QUANTERA European Union’s Horizon H2020 project "QUANTOX" under Grant Agreement No. 731473. 
M.C., P.G.\ and Y.F.\ acknowledge support by the project “Two-dimensional Oxides Platform for SPINorbitronics nanotechnology (TOPSPIN)” funded by the MIUR-PRIN Bando 2017 - grant 20177SL7HC.
M.C. acknowledges support by the EU’s Horizon 2020 research and innovation program under Grant Agreement nr.\ 964398 (SUPERGATE).
This work is supported by the JSPS KAKENHI (Grants No.\ JP18H01176, No.\ JP18K03538, No.\ JP20H00131, and
No.\ JP20H01857) from MEXT of Japan,
Researcher Exchange Program between JSPS and RFBR (Grant No. JPJSBP120194816), and the JSPS Core-to-Core program Oxide
Superspin international network (Grants No.\ JPJSCCA20170002).
We acknowledge valuable discussions with C.\ Guarcello and M.T.\ Mercaldo.

Correspondence should be addressed to Mario Cuoco, email: mario.cuoco@spin.cnr.it.


\appendix


\section{Model}

The Hamiltonian in the normal state $\hat{H}(\bm{k})$ is given by
\begin{align}
    \hat{H}(\bm{k})=\hat{H}_{0}(\bm{k})+\hat{H}_\mathrm{SO}+\hat{H}_\mathrm{is}(\bm{k}).
\end{align}%
where $\hat{H}_0(\bm{k})$ denotes the kinetic term,
\begin{align}
    \hat{H}(\bm{k})=\hat{\varepsilon}(\bm{k})\otimes \hat{\sigma}_{0},
\end{align}%
\begin{align}
    \hat{\varepsilon}(\bm{k})&=
    \begin{pmatrix}
        \varepsilon_{yz}(\bm{k}) & 0 & 0 \\
        0 & \varepsilon_{zx}(\bm{k}) & 0 \\
        0 & 0 & \varepsilon_{xy}(\bm{k})
    \end{pmatrix},\\
    \varepsilon_{yz}(\bm{k})&=2t_1(1-\cos{k_y})+2t_3(1-\cos{k_x}),\\
    \varepsilon_{zx}(\bm{k})&=2t_1(1-\cos{k_x})+2t_3(1-\cos{k_y}),\\
    \varepsilon_{xy}(\bm{k})&=4t_2-2t_2(\cos{k_x}+\cos{k_y})+\Delta_\mathrm{t}.
\end{align}%
$\hat{H}_\mathrm{SO}$ and $\hat{H}_\mathrm{is}(\bm{k}$) stand for the atomic spin-orbit coupling and inversion symmetry breaking terms,
\begin{align}
    \hat{H}_\mathrm{SO}&=\lambda_\mathrm{SO}
    \hat{\bm{L}}\cdot\hat{\bm{\sigma}},\\
    \hat{H}_\mathrm{is}(\bm{k})&=\alpha_\mathrm{OR}[\hat{L}_{y}\sin{k_x}-\hat{L}_{x}\sin{k_y}],    
\end{align}%
respectively.
Here, $\hat{\sigma}_{i=0,x,y,z}$ denote the Pauli matrices in spin space and $\hat{L}_{j=0,x,y,z}$ the $t_{2g}$-orbital angular momentum operators projected onto $L=2$,
\begin{align}
    \hat{L}_x&=
    \begin{pmatrix}
        0 & 0 & 0 \\
        0 & 0 & i \\
        0 & -i & 0
    \end{pmatrix},\\
    \hat{L}_y&=
    \begin{pmatrix}        
        0 & 0 & -i \\
        0 & 0 & 0 \\
        i & 0 & 0
    \end{pmatrix},\\
    \hat{L}_z&=
    \begin{pmatrix}
        0 & i & 0 \\
        -i & 0 & 0 \\
        0 & 0 & 0
    \end{pmatrix},
\end{align}%
in the $[d_{yz},d_{zx},d_{xy}]$ basis.

For the (100) oriented surface, the local term $\tilde{u}(k_y)$ and nearest neighbor hopping matrix $\tilde{t}(k_y)$ can be explicitly derived.
The local term $\tilde{u}(k_y)$ is given by
\begin{align}
    \tilde{u}(k_y)&=
    \begin{pmatrix}
        \hat{u}(k_y) & \hat{\Delta}_\mathrm{B1} \\
        \hat{\Delta}^{\dagger}_\mathrm{B1} & -\hat{u}^{*}(-k_y)
    \end{pmatrix},
\end{align}%
\begin{align}
    \hat{u}(k_y)&=\hat{u}_{0}(k_y)+\hat{u}_\mathrm{SO}+\hat{u}_\mathrm{is}(k_y),\\
    \hat{u}_{0}(k_y)&=
    \begin{pmatrix}
        \tilde{\varepsilon}_{yz}(k_y) & 0 & 0 \\
        0 & \tilde{\varepsilon}_{zx}(k_y) & 0 \\
        0 & 0 & \tilde{\varepsilon}_{xy}(k_y)
    \end{pmatrix}
    \otimes\hat{\sigma}_{0},
\end{align}%
\begin{align}
    \tilde{\varepsilon}_{yz}(k_y)&=2t_1(1-\cos{k_y})+2t_3,\\
    \tilde{\varepsilon}_{zx}(k_y)&=2t_1+2t_3(1-\cos{k_y}),\\
    \tilde{\varepsilon}_{xy}(k_y)&=4t_2-2t_2\cos{k_y}+\Delta_\mathrm{t},
\end{align}%
\begin{align}
    \hat{u}_\mathrm{SO}&=\hat{H}_\mathrm{SO},\\
    \hat{u}_\mathrm{is}(k_y)&=-\alpha_\mathrm{OR}\hat{L}_{x}\otimes\hat{\sigma_{0}}\sin{k_y},
\end{align}%
and the nearest neighbor hopping $\tilde{t}(k_y)$,
\begin{align}
    \tilde{t}(k_y)=
    \begin{pmatrix}
        \hat{t}(k_y) & 0 \\
        0 & -\hat{t}^{*}(-k_y)
    \end{pmatrix},
\end{align}%
\begin{align}
    \hat{t}(k_y)=
    \begin{pmatrix}
        -t_3 & 0 & \frac{\alpha_\mathrm{OR}}{2} \\
        0 & -t_1 & 0 \\
        -\frac{\alpha_\mathrm{OR}}{2} & 0 & -t_2
    \end{pmatrix}
    \otimes\hat{\sigma_{0}}.
\end{align}%


\section{Recursive Green function method}

In the superconducting state the Green's function is expressed as
\begin{align}
    \tilde{G}(\bm{k},i\varepsilon_n)=\frac{1}{i\varepsilon_n-\hat{H}_\mathrm{BdG}(\bm{k})}
    =\left(\begin{array}{cc}
        \hat{G}(\bm{k},i\varepsilon_n) & \hat{F}(\bm{k},i\varepsilon_n) \\
        \bar{F}(\bm{k},i\varepsilon_n) & \bar{G}(\bm{k},i\varepsilon_n) 
    \end{array}\right),
\end{align}%
with the anomalous Green's function for the orbital indices $(\alpha,\beta)$ which is given by
\begin{align}
    \hat{F}^{(\alpha,\beta)}(\bm{k},i\varepsilon_n)=
    \begin{pmatrix}
        F^{(\alpha,\beta)}_{\uparrow \uparrow}(\bm{k},i\varepsilon_n) & F^{(\alpha,\beta)}_{\uparrow \downarrow}(\bm{k},i\varepsilon_n) \\
        F^{(\alpha,\beta)}_{\downarrow \uparrow}(\bm{k},i\varepsilon_n) & F^{(\alpha,\beta)}_{\downarrow \downarrow}(\bm{k},i\varepsilon_n)
    \end{pmatrix}.
\end{align}%
Since we consider the interorbital even-frequency/spin-triplet/orbital-singlet/$s$-wave pair potential, we focus on the \textbf{d}-vector in this pair amplitude.
We define the \textbf{d}-vector for the interorbital even-frequency/spin-triplet/orbital-singlet/$s$-wave pair amplitude,
\begin{align}
    \bm{d}^{(\alpha,\beta)}=(d^{(\alpha,\beta)}_{x},d^{(\alpha,\beta)}_{y},d^{(\alpha,\beta)}_{z}),
\end{align}
\begin{align}
    d^{(\alpha,\beta)}_{x}&=\frac{1}{2}\left[ F^{(\alpha,\beta)}_{\downarrow \downarrow}-F^{(\alpha,\beta)}_{\uparrow \uparrow} \right],\\
    d^{(\alpha,\beta)}_{y}&=\frac{1}{2i}\left[ F^{(\alpha,\beta)}_{\uparrow \downarrow}+F^{(\alpha,\beta)}_{\downarrow \downarrow} \right],\\
    d^{(\alpha,\beta)}_{z}&=F^{(\alpha,\beta)}_{\uparrow \downarrow+\downarrow \uparrow},
\end{align}%
Then, we also define
\begin{align}
    \bm{d}^{(j)}&=(d^{(j)}_x,d^{(j)}_y,d^{(j)}_z),
\end{align}%
\begin{align}
    d^{(j)}_x&=d^{(xy,yz)}_{xj}+d^{(xy,zx)}_{xj},\\
    d^{(j)}_y&=d^{(xy,yz)}_{yj}+d^{(xy,zx)}_{yj},\\
    d^{(j)}_z&=d^{(xy,yz)}_{zj}+d^{(xy,zx)}_{zj},
\end{align}
with $j=L,R$ superconductors, respectively.
In the numerical calculation, both $\bm{d}^{(L)}$ and $\bm{d}^{(R)}$ are real number.
Thus, the misalignment of \textbf{d}-vectors between two superconductors is obtained by
\begin{align}
    \gamma(a,b)=\theta_\mathrm{R}(b)-\theta_\mathrm{L}(a),
\end{align}%
with $\theta_{j}(a)=\arg[d^{(j)}_x+id^{(j)}_{y}]$ for the band indices $a$ and $b$.


Here, we report the main steps for the determination of the Josephson current by the recursive Green's function method \cite{LDOSUmerski}.

\begin{align}
    \tilde{u}_\mathrm{L}(k_y)&=-\mu_\mathrm{L}\hat{L}_{0}\otimes\hat{\sigma}_{0}+\tilde{u}(k_y),\\
    \tilde{t}_\mathrm{L}&=\tilde{t}(k_y),
\end{align}%
\begin{align}
    \tilde{u}_\mathrm{R}(k_y)=-\mu_\mathrm{R}\hat{L}_{0}\otimes\hat{\sigma}_{0}+\tilde{u}(k_y),
\end{align}%
\begin{align}
    \tilde{t}_\mathrm{R}=\tilde{t}(k_y),
\end{align}%
\begin{align}
    \tilde{u}_\mathrm{N}(k_y)=-\mu_\mathrm{N}\hat{L}_{0}\otimes\hat{\sigma}_{0}+
    \begin{pmatrix}
        \hat{u}(k_y) & 0 \\
        0 & -\hat{u}^{*}(-k_y)
    \end{pmatrix},
\end{align}%
\begin{align}
    \tilde{t}_\mathrm{N}=\tilde{t}(k_y),
\end{align}
with the chemical potentials in left (right)-side superconductors $\mu_\mathrm{L}$ ($\mu_\mathrm{R}$), and the normal layer $\mu_\mathrm{N}$, respectively.

\begin{align}
    \hat{t}_\mathrm{L0}
    =\hat{t}_\mathrm{1R}=t_\mathrm{int}\tilde{t}(k_y),
\end{align}
with the transparency $t_\mathrm{int}$.
In this calculation, we fix the transparency as $t_\mathrm{int}=1.0$.

First, we calculate the semi-infinite surface Green's function $\hat{G}_\mathrm{L}$ and $\hat{G}_\mathrm{R}$ in the left and right-side superconductors, respectively\cite{LDOSUmerski}.
When we add a normal layer, we obtain the surface Green's functions $\hat{G}_\mathrm{L0}(k_y,i\varepsilon_{n})$ and $\hat{G}_\mathrm{R1}(k_y,i\varepsilon_{n},\phi)$,
\begin{align}
    \hat{G}_\mathrm{L0}(k_y,i\varepsilon_{n})&=[i\varepsilon_{n}-\tilde{u}_\mathrm{N}-\hat{t}^{\dagger}_\mathrm{L0}\hat{G}_\mathrm{L}\hat{t}_\mathrm{L0}]^{-1}, \\
    \hat{G}_\mathrm{R1}(k_y,i\varepsilon_{n},\phi)&=[i\varepsilon_{n}-\tilde{u}_\mathrm{N}-\hat{t}_\mathrm{1R}\hat{G}_\mathrm{R}\hat{t}^{\dagger}_\mathrm{1R}]^{-1},
\end{align}%
with the local term in the normal layer $\tilde{u}_\mathrm{N}$, the tunnel Hamiltonian $\hat{t}_\mathrm{L0}=\hat{t}_\mathrm{1R}$, and the fermionic Matsubara frequency $i\varepsilon_n=i\pi(2n+1)k_\mathrm{B}T$.
For the connection of two superconductors with a normal layer, we calculate the local Green's functions,
\begin{align}
    \hat{G}_{00}(k_y,i\varepsilon_{n},\phi)&=[\hat{G}^{-1}_\mathrm{L0}-\tilde{t}_\mathrm{N}\hat{G}_\mathrm{R1}\tilde{t}^{\dagger}_\mathrm{N}]^{-1}, \\
    \hat{G}_{11}(k_y,i\varepsilon_{n},\phi)&=[\hat{G}^{-1}_\mathrm{R1}-\tilde{t}^{\dagger}_\mathrm{N}\hat{G}_\mathrm{L0}\tilde{t}_\mathrm{N}]^{-1},
\end{align}%
and the nonlocal Green's functions,
\begin{align}
    \hat{G}_{01}(k_y,i\varepsilon_{n},\phi)&=\hat{G}_\mathrm{L0}\tilde{t}_\mathrm{N}\hat{G}_{11}, \\
    \hat{G}_{10}(k_y,i\varepsilon_{n},\phi)&=\hat{G}_\mathrm{R1}\tilde{t}^{\dagger}_\mathrm{N}\hat{G}_{00}.
\end{align}%






\end{document}


\title{Supplemental Material: Anomalous Josephson Coupling and High-Harmonics in Non-Centrosymmetric Superconductors with $S$-wave Spin-Triplet Pairing}

\author{Yuri Fukaya}
\affiliation{CNR-SPIN, I-84084 Fisciano (Salerno), Italy, c/o Universit\'a di Salerno, I-84084 Fisciano (Salerno), Italy}

\author{Yukio Tanaka}
\affiliation{Department of Applied Physics, Nagoya University, Nagoya 464-8603, Japan}
\author{Paola Gentile}
\affiliation{CNR-SPIN, I-84084 Fisciano (Salerno), Italy, c/o Universit\'a di Salerno, I-84084 Fisciano (Salerno), Italy}

\author{Keiji Yada}
\affiliation{Department of Applied Physics, Nagoya University, Nagoya 464-8603, Japan}

\author{Mario Cuoco}
\affiliation{CNR-SPIN, I-84084 Fisciano (Salerno), Italy, c/o Universit\'a di Salerno, I-84084 Fisciano (Salerno), Italy}

\maketitle

\section{Josephson current}

\begin{figure}[htbp]
    \centering
    \includegraphics[width=16cm]{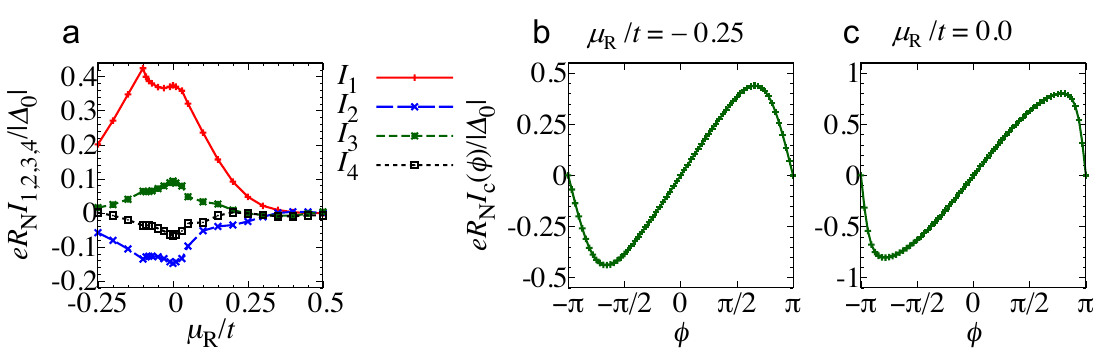}
    \caption{\textbf{Harmonics of the Josephson current and current-phase relation for different chemical potential.} (a) Amplitude of the first four harmonics of the Josephson current $I_{n=1,2,3,4}$ as a function of the chemical potential on the right-side superconductor $\mu_\mathrm{R}$
    at $(\lambda_\mathrm{SOL}/t,\alpha_\mathrm{ORL}/t,\mu_\mathrm{L}/t)=(0.10,0.20,0.00)$, $(\lambda_\mathrm{SOR}/t,\alpha_\mathrm{ORR}/t)=(0.10,0.20)$ and $T/T_\mathrm{c}=0.10$. 
    (b,c) Corresponding current phase relations in (a) at (b) $\mu_\mathrm{R}/t=-0.25$ and (c) $\mu_\mathrm{R}/t=0.0$. 
    We showed the current phase relation at $\mu_\mathrm{R}/t=0.35$ in Fig.\ 3 (e) of the main part.}
    \label{SMfig1}
\end{figure}
\begin{figure}[htbp]
    \centering
    \includegraphics[width=14cm]{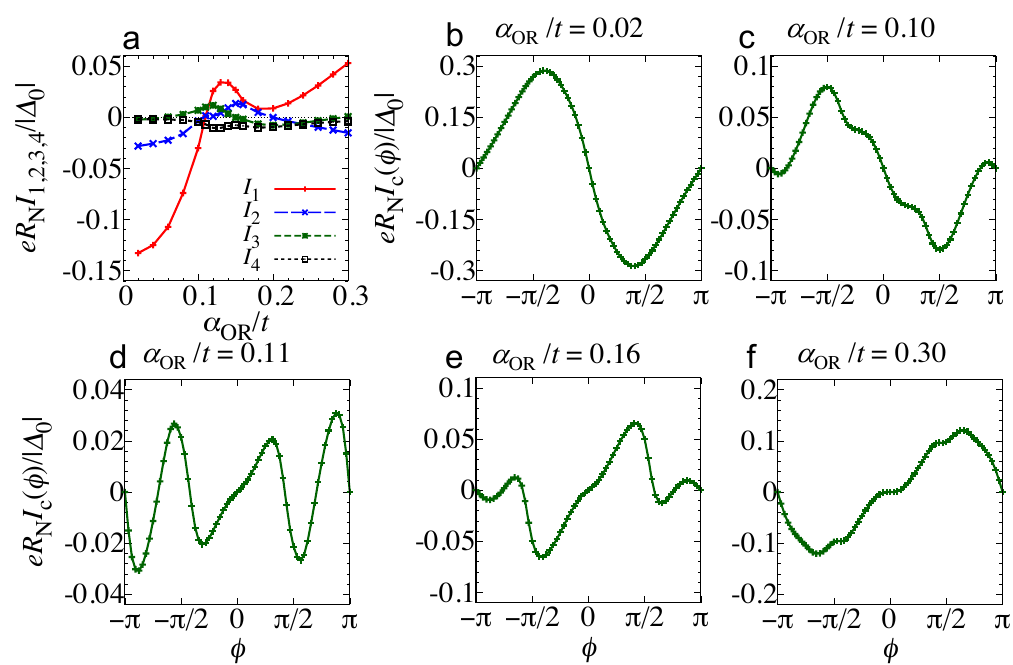}
    \caption{\textbf{Harmonics of the Josephson current and current-phase relation for different values of the orbital Rashba interaction.} (a) Amplitude of the first four harmonics of normalized Josephson current, $I_{n=1,2,3,4}$, as a function of $\alpha_\mathrm{OR}=\alpha_\mathrm{ORL}=\alpha_\mathrm{ORR}$.
    (b,c,d,e,f) Corresponding current phase relations in (a) at (b) $\alpha_\mathrm{OR}/t=0.02$, (c) $\alpha_\mathrm{OR}/t=0.10$, (d) $\alpha_\mathrm{OR}/t=0.11$,(e) $\alpha_\mathrm{OR}/t=0.16$, and (f) $\alpha_\mathrm{OR}/t=0.30$.
    We set the parameters as $(\lambda_\mathrm{SOL}/t,\mu_\mathrm{L}/t)=(0.10,0.00)$,  $(\lambda_\mathrm{SOR}/t,\mu_\mathrm{R}/t)=(0.10,0.35)$, and $T/T_\mathrm{c}=0.025$. 
    }
    \label{SMfig2}
\end{figure}
\begin{figure}[htbp]
    \centering
    \includegraphics[width=17cm]{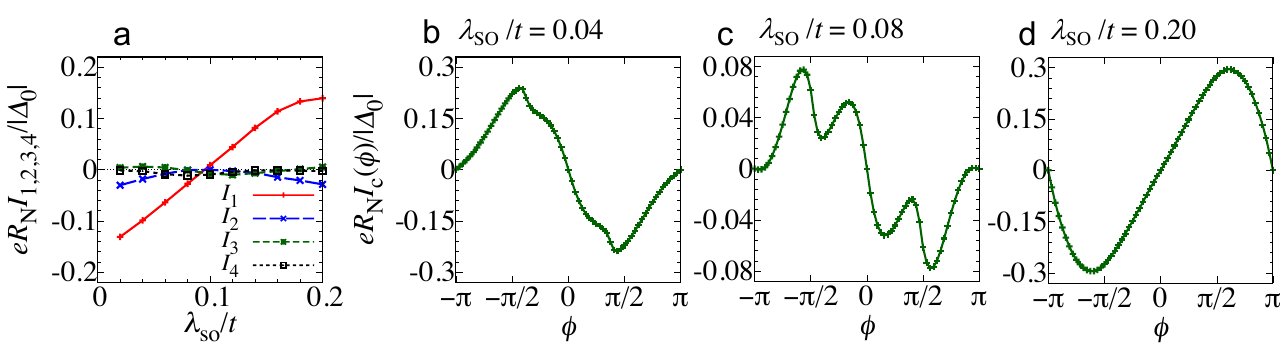}
    \caption{\textbf{Harmonics of the Josephson current and current-phase relation for different values of the atomic spin-orbit coupling.} (a) Components of normalized Josephson current  $I_{n=1,2,3,4}$ as a function of $\lambda_\mathrm{SO}=\lambda_\mathrm{SOL}=\lambda_\mathrm{SOR}$.
    (b,c,d) Corresponding current phase relations in (a) at (b) $\lambda_\mathrm{SO}/t=0.04$, (c) $\lambda_\mathrm{SO}/t=0.08$, and (d) $\lambda_\mathrm{SO}/t=0.20$.
    We set the parameters as $(\alpha_\mathrm{ORL}/t,\mu_\mathrm{L}/t)=(0.20,0.00)$,  $(\alpha_\mathrm{ORR}/t,\mu_\mathrm{R}/t)=(0.20,0.35)$, and $T/T_\mathrm{c}=0.025$. 
    }
    \label{SMfig3}
\end{figure}

We report about the evolution of the Josephson current as a function of the electronic parameters. We start by considering a configuration with an amplitude mismatch of the chemical potentials between the left ($\mu_\mathrm{L}$) and right ($\mu_\mathrm{R}$) side of the junction (Fig. \ref{SMfig1}). We track the evolution of the first four harmonics of the Josephson current as a function of $\mu_R$ at a given value of $\mu_\mathrm{L}=0.0$ (Fig. \ref{SMfig1}(a) as well of the CPR for two representative cases of $\mu_\mathrm{L}$ ((Fig. \ref{SMfig1}(b),(c)). For these configurations, corresponding to equal amplitude of the spin-orbit ($\lambda_\mathrm{SO}$) and orbital Rashba couplings on the two sides of the junction, the CPR is marked by a 0-type Josephson coupling.
It is more interesting to observe the evolution as a function of the spin-orbit and orbital Rashba interactions assuming a fixed amplitude mismatch of the chemical potentials. In Figs. \ref{SMfig2},\ref{SMfig3} we show the evolution of the first four harmonics and the corresponding CPR as a function of the orbital Rashba coupling ($\alpha_{\text{OR}}$) and the spin-orbit coupling ($\lambda_{\text{SO}}$). We find that the first harmonic can change sign at a critical value of $\alpha_{\text{OR}}$ or $\lambda_{\text{SO}}$ thus signalling a $0$- to $\pi$-phase transition. In proximity of the transition the CPR is dominated by high harmonics as manifested by the presence of double maximum and asymmetric amplitudes of the maximum and minimum in the current phase relations.


To evaluate the dependence of the critical current on an applied magnetic field $H$ we follow the standard approach for a short junction with length $L$ and width $W$ that is thread by a magnetic flux $\Phi= L W H$. 
Then, we can expand the current phase relation in all the harmonics and consider that the spatial dependent vector potential associated to the magnetic field modifies the phase difference $\phi$ across the junction with a factor $2\pi\tilde{\Phi}\tilde{x}$ where $\tilde{\Phi}=\Phi/\Phi_0$, with $\Phi_0$ being the unit of magnetic flux, and $\tilde{x}$ indicating the ratio $x/W$, i.e. the scaled length with respect to the lateral size of the junction.
The general expression of the supercurrent expressed via an expansion in harmonics is given by
\begin{align}
    I(\tilde{\Phi},\phi)=\sum_{n}\int^{w}_{0}I_{n}\sin\left[n \phi+2\pi\tilde{\Phi}\tilde{x}\right]d\tilde{x} \,.
\end{align}%

The resulting critical current at a given applied magnetic flux is obtained by maximizing $I(\tilde{\Phi},\phi)$ with respect to $\phi$.

